\documentclass[twocolumn,preprintnumbers,amsmath,amssymb]{revtex4}
\usepackage{color}
\usepackage{mathptmx}
\usepackage{amsmath} 
\usepackage{amssymb} 
\usepackage{graphicx}% Include figure files
\usepackage{dcolumn}% Align table columns on decimal point
\usepackage{bm}% bold math
\newcommand{\mean}[1]{\langle#1\rangle}
\newcommand{\ket}[1]{\mid#1\rangle}
\newcommand{\bra}[1]{\langle#1\mid} 

\newcommand{\Tr}[1]{\mathrm{Tr}(#1)}

\begin{document}

%\begin{widetext}
%  \tableofcontents
%\end{widetext}

\preprint{APS/123-QED}

\title{Steady state entanglement of two coupled qubits}

\author{Elena del Valle} \email{elena.delvalle.reboul@gmail.com}
\affiliation{School of Physics and Astronomy, University of
  Southampton, SO17 1BJ, Southampton, United Kingdom}

\date{\today}% It is always \today, today,

\begin{abstract}
  The maximum entanglement between two coupled qubits in the steady
  state under two independent incoherent sources of excitation is
  reported. Asymmetric configurations where one qubit is excited while
  the other one dissipates the excitation are optimal for
  entanglement, reaching values three times larger than with thermal
  sources. The reason is the purification of the steady state mixture
  (that includes a Bell state) thanks to the saturation of the pumped
  qubit. Photon antibunching between the cross emission of the qubits
  can be used to experimentally evidence the large degrees of
  entanglement.
%PACS numbers may be entered using the \verb+\pacs{#1}+ command.
\end{abstract}

\maketitle
\section{Introduction}

A global state of a composite system is entangled when it cannot be
written as a product of the states of the individual
systems~\cite{horodecki09a}. This is the basic quantum-mechanical
property, with no classical analog, for quantum information
technologies~\cite{nielsen_book00a,ladd10a}. Two coupled qubits, or
two-level systems with ground $\ket{g}$ and excited $\ket{e}$ states,
is the smallest and simplest composite system that can display
entanglement. It is, therefore, the most suitable model to investigate
its creation and processing as well as how environmental noise and
decoherence brought by spontaneous decay and the external excitation
affects it~\cite{gardiner_book00a}, which is a key point for quantum
applications.

Two qubits can form four independent maximally entangled states, the
so-called \emph{Bell states}: $|\phi_\pm\rangle=(|gg\rangle\pm
|ee\rangle)/\sqrt{2}$ and $|\psi_\pm\rangle=(|eg\rangle\pm
|ge\rangle)/\sqrt{2}$. The last two are affected by a possible
coupling between the qubits, and are also known in the atomic
literature as the \emph{symmetric} and \emph{antisymmetric collective
  states}~\cite{ficek02a}. The formation and degradation of such
states when subjected to spontaneous emission has been the object of
much recent research~\cite{verstraete09a,yu09a,ficek10a}, focusing on
the preservation of entanglement into \emph{decoherence-free
  subspaces} and taking advantage of the collective damping or
effective coupling created between the qubits by interaction with
common
reservoirs~\cite{braun02a,kim02a,jakobczyk02a,schneider02a,benatti03a,xiang-ping06a,delvalle07a,delvalle07b,contreras-pulido08a,hor-meyll09a,benatti10a,jakobczyk10a}.

The idea of environmentally induced entanglement has also been applied
to the case where the two qubit interaction is mediated by a cavity
mode (harmonic oscillator) which is excited by white noise (a thermal
reservoir)~\cite{yi03a}, borrowing the idea from Ref.~\cite{plenio02a}
where, on the contrary, entanglement is enhanced between two harmonic
modes by mediation of a two-level system excited by white noise. In
both cases, entanglement may survive in a steady state that is not the
vacuum, but is very small ($<0.4\%$). A two-level system has also been
proposed as a mediator (or \emph{coupler}), to build entanglement
between qubits~\cite{hao07a,huang09a}.

Another possibility, close to the one addressed in the present text,
is to consider two qubits already coupled, whose entanglement builds
in the steady state despite dissipation and decoherence from two
\emph{independent}
environments~\cite{xujb05a,hartmann06a,hartmann07a,lambert07a,rivas09a,wang09a,zhou09a,li09b,shan10a}. Entanglement,
being essentially a property that requires great purity of the state,
is very sensitive to such decoherence. It is therefore important to
look for optimization. Having a high degree of entanglement in the
steady state means that it is robust and independent of the initial
state, it remains stored forever in our system like a \emph{quantum
  battery of entanglement}~\cite{li09b}.

The coupling between the qubits can have different physical origins
depending on the realization~\cite{ladd10a}, e.g., Rydberg atoms
couple through dipole-dipole interaction, weaker in the case of cold
atoms~\cite{bloch08b}, and so do excitons in single quantum dots or
molecules~\cite{bayer01b,krenner05b,gerardot05a}; superconducting
qubits couple through mutual inductance~\cite{clarke08a}. Moreover, in
all these implementations, the coupling can appear effectively through
the virtual mediation of a coupler (a cavity or a wire mode) in the
dispersive limit, in which case it is given by $g_\mathrm{eff}\approx
G_\mathrm{c}^2/\Delta_\mathrm{c}$, where $G_\mathrm{c}$ is the
coupling of the qubits to the coupler and $\Delta_\mathrm{c}$ the
energy detuning to the qubits (considered much larger than the
coupling)~\cite{imamoglu99a,shi-biao00a,ashhab08a}. This scheme
requires, for instance, placing the qubits into a cavity where the
cavity mode acts as the coupler. One can take advantage of the QED
techniques while obtaining an effective coupling essentially
insensitive to the cavity decay and thermal fluctuations. The
effective coupling between two Rydberg atoms through virtual photon
exchange, while crossing a nonresonant cavity, was achieved in
2001~\cite{osnaghi01a}. The final entangled state could be controlled
by adjusting the atom-cavity detuning. A similar effective coupling
was obtained between two superconducting qubits on opposite sides of a
chip using microwave photons confined in a transmission line
cavity~\cite{majer07a}. The cavity was also used to perform
multiplexed control and measurement of both qubit states. Effective
coupling between two distant quantum dots embedded in a microcavity
has also been recently achieved~\cite{arxiv_laucht10a,gallardo10a}.

Taking for granted that the two qubits are coupled, we center our
attention on the situation where the qubits are also in contact with
two independent excitation sources. Xu and Li~\cite{xujb05a} found
that with two equally intense white-noise sources at the same
temperature, no entanglement can be formed in the steady
state. However, if only one qubit was subjected to a finite
temperature source, some entanglement could be achieved. They also
pointed out that the steady state entanglement exhibits a typical
double \emph{stochastic-resonance} as a function of the decoherence
parameters of both qubits~\cite{plenio02a}. They found better but
still small degrees of entanglement ($<4\%$) and did not deepen on its
origin, but their results show that an asymmetric flow of excitation
through the qubits is beneficial for entanglement. Other authors, who
did not consider unequal sources of excitation (having to recur to
other mechanisms for entanglement generation), explored, on the other
hand, configurations that are out of thermal equilibrium, where the
excitation of the qubits is not necessarily produced by thermal
sources but by more general processes able of inverting the qubits
population~\cite{hartmann07a,lambert07a,zhou09a}. Two qubits may
undergo dissipation and pure dephasing but also an externally
controllable and independent (in general) continuous pumping that can
have great impact on the strong coupling reached in the steady
state~\cite{delvalle_book10a,delvalle10b}.

In the present text, I put together different elements that have been
addressed separately in previous studies of environmentally induced
entanglement: direct coupling between the qubits and independent and
different kinds of reservoirs that are not necessarily of a thermal
nature. I give a complete picture of entanglement and its origin in
the steady state of such a general system. As a result, I find a
configuration where entanglement is significantly enhanced ($31\%$),
that is, more than three times as compared to the best thermal case
and with a much better purity. I show that it can be evidenced by the
antibunching of the two qubits cross emission.

The rest of this paper is organized as follows. In
Sec.~\ref{sec:MonMay17124011BST2010}, I introduce a theoretical model
to describe two coupled qubits with decay, incoherent pumping and pure
dephasing, and a quantity to quantify the degree of entanglement
between them, the concurrence. In
Sec.~\ref{sec:MonMay17124037BST2010}, I discuss different entangled
configurations and optimize the concurrence for the most suitable one:
one qubit is excited while the other dissipates the excitation. This
is compared with the thermal counterpart. In
Sec.~\ref{sec:MonMay17124051BST2010}, I show how a strong antibunching
between the two qubits emissions, is linked with high degrees of
entanglement and I propose this effect as an indication of
entanglement. In Sec.~\ref{sec:MonMay17124101BST2010}, I study the
effect of pure dephasing on entanglement. In
Sec.~\ref{sec:MonMay17124110BST2010}, I present the conclusions.

\section{Theoretical model}
\label{sec:MonMay17124011BST2010}

\begin{figure}[t]
\centering
\includegraphics[width=0.9\linewidth]{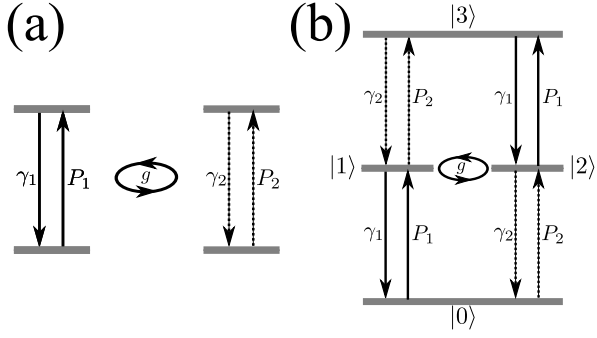}
\caption{Scheme of the two coupled qubits or two-level systems (a) and
  their energy levels (b), with coupling ($g$), pumping ($P_i$) and
  decay parameters ($\gamma_i$).}
\label{fig:FriJan23211556GMT2009}
\end{figure}

Let us consider two qubits or two-level systems ($i=1,2$), with
lowering operators~$\sigma_i$, frequencies~$\omega_i$ and coupled with
strength~$g$. Without loss of generality, we take the energy of the
first qubit as a reference ($\omega_1=0$), from which the other one is
detuned by a small quantity $\Delta=\omega_1-\omega_2$. The
corresponding Hamiltonian reads:
\begin{equation}
  \label{eq:SunDec21163832CET2008}
  H=-\Delta \sigma_2^\dagger\sigma_2+g(\sigma_1^\dagger\sigma_2+\sigma_2^\dagger\sigma_1)\,,
\end{equation}
Each qubit can be in the ground ($|g\rangle$) or excited ($|e\rangle$)
states, whose direct product produces a Hilbert space of dimension $4$
(see Fig.~\ref{fig:FriJan23211556GMT2009}): $\{|0\rangle=|gg\rangle$,
$|1\rangle=|eg\rangle$, $|2\rangle=|ge\rangle$,
$|3\rangle=|ee\rangle\}$.  The qubits are in contact with different
kinds of environments, that provide or dissipate excitation (at rates
$P_i$ and $\gamma_i$, respectively) in an incoherent continuous
way. Other interactions may purely bring dephasing to the coherent
dynamics (at rates $\gamma_i^d$). These processes eventually drive any
pure state into a statistical mixture of all possible states. A
density matrix, $\rho$, properly describes the evolution of such a
system. The general master equation we consider has the standard
Liouvillian form~\cite{carmichael_book02a}:
\begin{equation}
  \label{eq:ME}
  \partial_t  \rho=i[\rho,H]+\sum_{i=1}^2\Big[\frac{\gamma_i}{2}\mathcal{L}_{\sigma_i} +\frac{P_i}{2}\mathcal{L}_{\sigma_i^\dagger}+\frac{\gamma_i^d}{2}\mathcal{L}_{\sigma_i^\dagger \sigma_i}\Big]\rho \,,
\end{equation}
with the corresponding Lindblad terms for the incoherent processes
($\mathcal{L}_{O}\rho\equiv 2 O \rho O^\dagger -O^\dagger O \rho- \rho
O^\dagger O$). If the two qubits shared a common environment, the
Lindblad terms would share a single expression $\mathcal{L}_{J}$ in
terms of the collective operator: $J=\sigma_1+\sigma_2$. Such
collective terms are sources of entanglement, as explained in the
introduction. In this text, I investigate the steady state of a system
where they are not present, solving exactly the equation~$\partial_t
\rho=0$.

In order to spell out the nature of the reservoirs that are in contact
with the qubits, I express the pumping and decay rates in terms of new
parameters $\Gamma_i$ and $r_i$~\cite{briegel93a}:
\begin{equation}
  \label{eq:TueOct13194019GMT2009}
  \gamma_i=\Gamma_i(1-r_i)\,,\quad P_i=\Gamma_i r_i\, \quad(i=1,2).
\end{equation}
The range $0\leq r_i \leq 1$, includes a medium that only absorbs
excitation (decay, $r_i=0$), one which only provides it (pump,
$r_i=1$), as well as a the most common assumption of a thermal bath
with finite temperature, (white noise, $r_i<1/2$). The parameter
$\Gamma_i=\gamma_i+P_i$ quantifies the interaction of each qubit with
its reservoir as well as its effective spectral broadening.

Taking into account the evolution of the density matrix $\rho$ of our
bipartite system, one can conclude that in the steady state, it has
the general block diagonal form:
\begin{eqnarray}
  \label{eq:SatFeb6235837WET2010}
  \rho&=& \left( \begin{array}{cccc}
      \rho_{00} & 0           & 0        & 0 \\
      0        & \rho_{11}    & \rho_{12} & 0  \\
      0        & \rho^*_{12}  & \rho_{22} &  0 \\
      0        & 0           & 0         & \rho_{33}\end{array} \right)
\end{eqnarray}
with
\begin{subequations}
  \label{eq:ThuMar18180901WET2010}
  \begin{align}
  &\rho_{00} = 1-\mean{n_1}-\mean{n_2}+\mean{n_1 n_2}\,,\\
  &\rho_{ii}=\mean{n_i}-\mean{n_1 n_2}  \,,\quad i=1,2\,,\\
  &\rho_{12}= \mean{n_{12}}^*\,,  \\
  &\rho_{33}=\mean{n_1 n_2}\,, 
\end{align}
\end{subequations}
in terms of the operators $n_i=\sigma_i^\dagger\sigma_i$ and
$n_{12}=\sigma_1^\dagger\sigma_2$. The average value $\mean{n_i}$ is
the probability for qubit $i$ to be excited, regardless of the other
one.  $\mean{n_{12}}$ accounts for the population transfer between the
qubits and $\mean{n_1n_2}$ for their effective coupling, in the sense
that, independent qubits would lead
to~$\mean{n_1n_2}=\mean{n_1}\mean{n_2}$. The general expressions for
the steady state of two coupled qubits can be written as
$\mean{n_i}=P_i^\mathrm{eff}/\Gamma_i^\mathrm{eff}$,
$\mean{n_1n_2}=(P_1 \mean{n_2}+P_2\mean{n_1})/(\Gamma_1+\Gamma_2)$,
$\mean{n_{12}}=\frac{2g(\mean{n_1}-\mean{n_2})}{2\Delta+i\Gamma_\mathrm{tot}}$,
where the effective parameters $P_i^\mathrm{eff}=P_i+(P_1+P_2)X_i$ and
$\Gamma_i^\mathrm{eff}=\Gamma_i+(\Gamma_1+\Gamma_2)X_i$ are expressed
in terms of an effective coherent exchange factor $X_i\equiv
\frac{4g^2/\Gamma_{3i-1}}{\Gamma_\mathrm{tot}[1+(2\Delta/\Gamma_\mathrm{tot})^2]}$
related to the Purcell rates~\cite{delvalle10b}. $X_i$ quantifies how
efficiently the external inputs and outputs are distributed among the
qubits thanks to the coherent coupling and despite the total
decoherence,
$\Gamma_\mathrm{tot}=\Gamma_1+\Gamma_2+\gamma_1^d+\gamma_2^d$.

In Ref.~\cite{delvalle10b}, I showed that the steady state $\rho$ of
two coupled qubits is the same than that of a four-level system, that
is, the system depicted in Fig.~\ref{fig:FriJan23211556GMT2009}(b)
with no correspondence to two two-level systems in
Fig.~\ref{fig:FriJan23211556GMT2009}(a), but rather to a single
entity. This is the case of four single atomic levels or of a single
quantum dot that can host two excitons and form a biexciton state. The
results presented in the following sections are directly based on
$\rho$ or on averaged (single-time) quantities $\mean{O}$ computed as
$\Tr{\rho O}$ and, therefore, are also valid for a four-level system.

\section{Entanglement and linear entropy}
\label{sec:MonMay17124037BST2010}

Among the four Bell states, $|\phi_\pm\rangle$ and $\ket{\psi_\pm}$,
only the last two are achievable in the present configuration (since
$\rho_{03}=0$). Let us therefore write the most general entangled
state that can be achieved in this system as $\ket{\psi}\equiv
(\ket{1}+e^{i\beta}\ket{2})/\sqrt{2}$.  Logically, the larger the
probability to find the qubits in such state (the closer $\rho$ is to
$\ket{\psi}\bra{\psi}$), the larger is the degree of entanglement in
the mixture represented by $\rho$. In order to make this statement
mathematically precise, we can make explicit the entangled
contribution to $\rho$ by expressing it as
\begin{multline}
  \label{eq:WedMay19125228BST2010}
  \rho=\rho_{00}\ket{0}\bra{0}+\rho_{33}\ket{3}\bra{3}\\+R_{1}\ket{1}\bra{1}+R_{2}\ket{2}\bra{2}+R_{\psi}\ket{\psi}\bra{\psi}
\end{multline}
where $R_{1}=\rho_{11}-|\rho_{12}|$, $R_{2}=\rho_{22}-|\rho_{12}|$ and
$R_{\psi}=2|\rho_{12}|$. $R_i$ ($i=1,2,\psi$) are not probabilities
($R_1$, $R_2$ may be negative) but, when they are normalized as
\begin{equation}
  \label{eq:WedMay19130029BST2010}
  \tilde R_i=\frac{|R_i|}{\rho_{00}+\rho_{33}+|R_1|+|R_2|+R_{\psi}}\,,
\end{equation}
they represent the contribution of the pure states $\ket{i}$ to the
mixture where the entangled state has been identified and set
apart. In order to enhance entanglement, we must maximize $\tilde
R_{\psi}$ ($\rho_{12}$) while minimizing the populations $\rho_{00}$
and $\rho_{33}$, and the differences $\tilde R_1$ and $\tilde
R_2$. The non-entangled contributions can be put together in a single
expression to be minimized: $\tilde R=1- \tilde R_{\psi}$.

The degree of entanglement can be quantified by the \emph{concurrence}
($C$)~\cite{wootters98a}, which ranges from~0 (separable states) to~1
(maximally entangled states). It defined as
$C\equiv[\mathrm{max}\{0,\sqrt{\lambda_1}-
\sqrt{\lambda_2}-\sqrt{\lambda_3}- \sqrt{\lambda_4}\}]$, where
$\{\lambda_1, \lambda_2, \lambda_3, \lambda_4 \}$ are the eigenvalues
in decreasing order of the matrix $\rho T \rho^*T$, with $T$ being a
anti diagonal matrix with elements $\{ -1,1,1,-1\}$. The concurrence
in this system is given by
$C=2\mathrm{Max}[\{0,|\rho_{12}|-\sqrt{\rho_{00}\rho_{33}}\}]$, which
shows a threshold behaviour that we anticipated above: the coherence
between the intermediate states, $|\rho_{12}|$, must overcome the
population of the spurious states $\rho_{00}$, $\rho_{33}$. A related
important factor to build some concurrence, is the degree of purity in
the system~\cite{munro01a}. This is measured through the \emph{linear
  entropy}, $S_L\equiv\frac{4}{3}[1 - {\rm Tr} (\rho^2)]$, which is 0
for a pure state, and 1 for a maximally disordered state (where all
four states occur with the same probability~$1/4$).

Without loss of generality, we can analyze the entanglement and linear
entropy of our system in the steady state, by considering the
parameters $\Delta\geq 0$, on the one hand, $0\leq r_2 \leq r_1 \leq
1$ on the other hand, and arbitrary $\Gamma_1,\Gamma_2\geq 0$.  This
simply implies that we label as 2 the qubit that is in contact with
the medium which has the most dissipative nature. Let us ignore
dephasing effects for the moment (we bring them back in
Sec.~\ref{sec:MonMay17124101BST2010}).

\begin{figure}[h!]
\centering
\includegraphics[width=\linewidth]{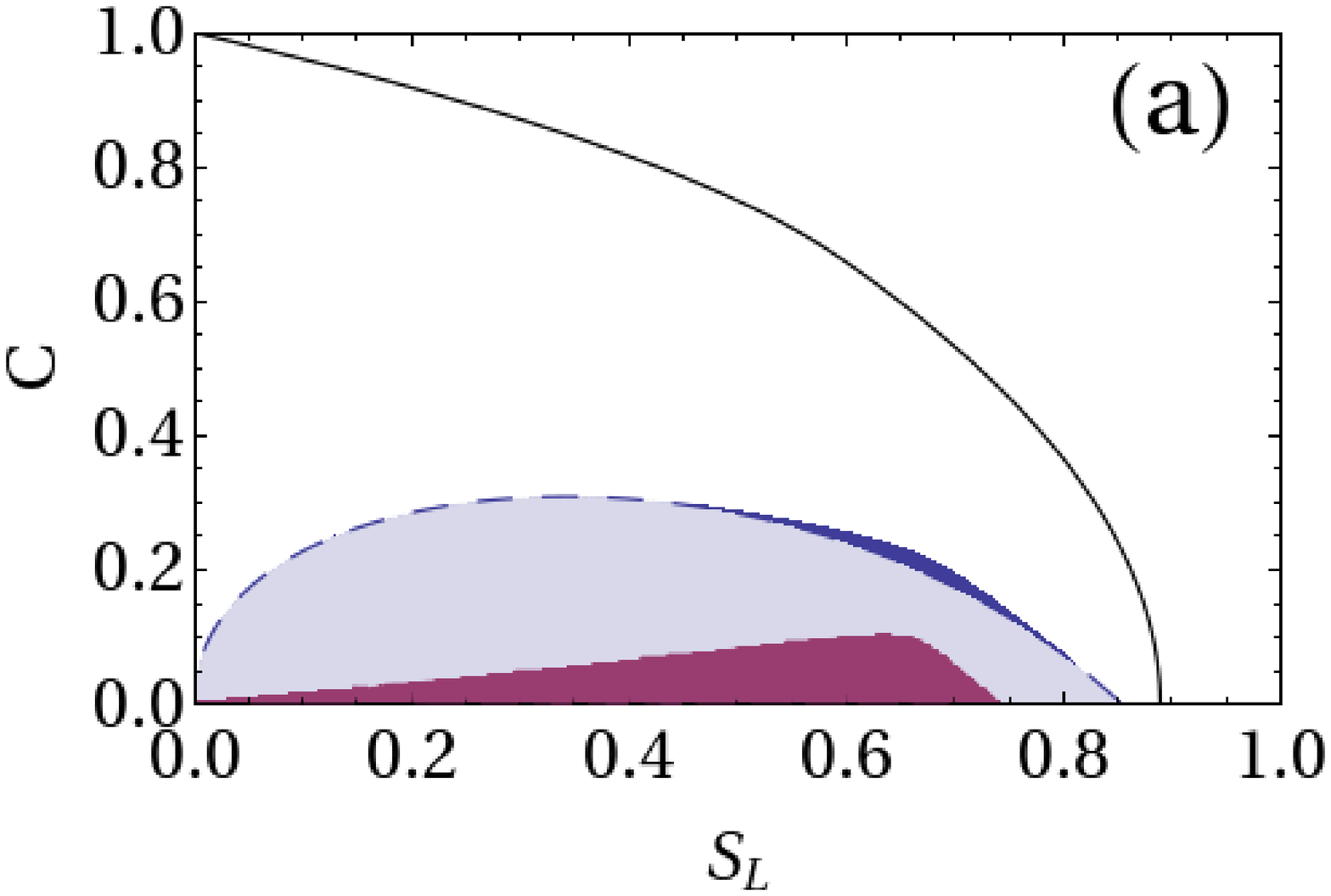}\\
\includegraphics[width=0.8\linewidth]{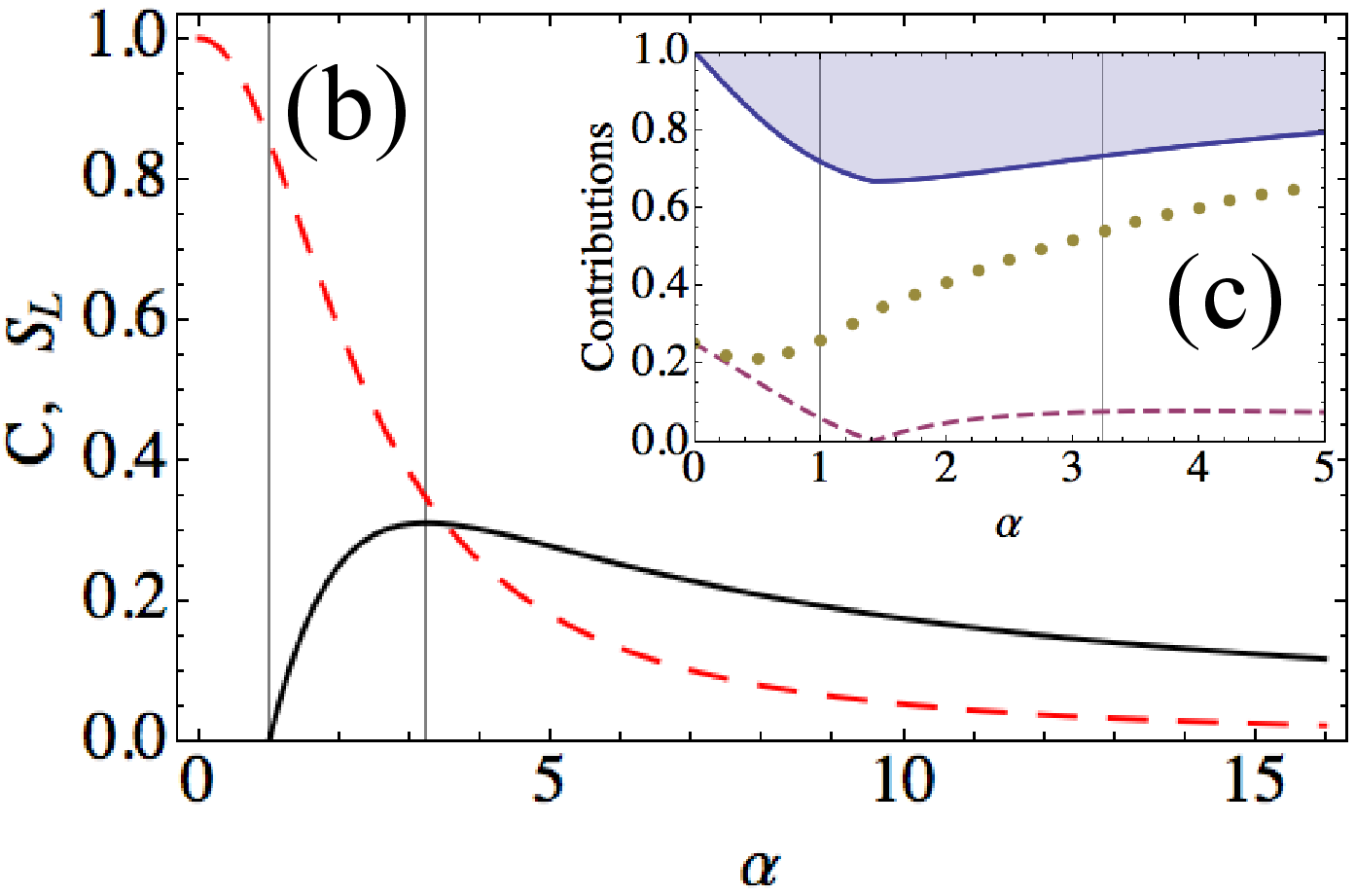}\\
\includegraphics[width=0.8\linewidth]{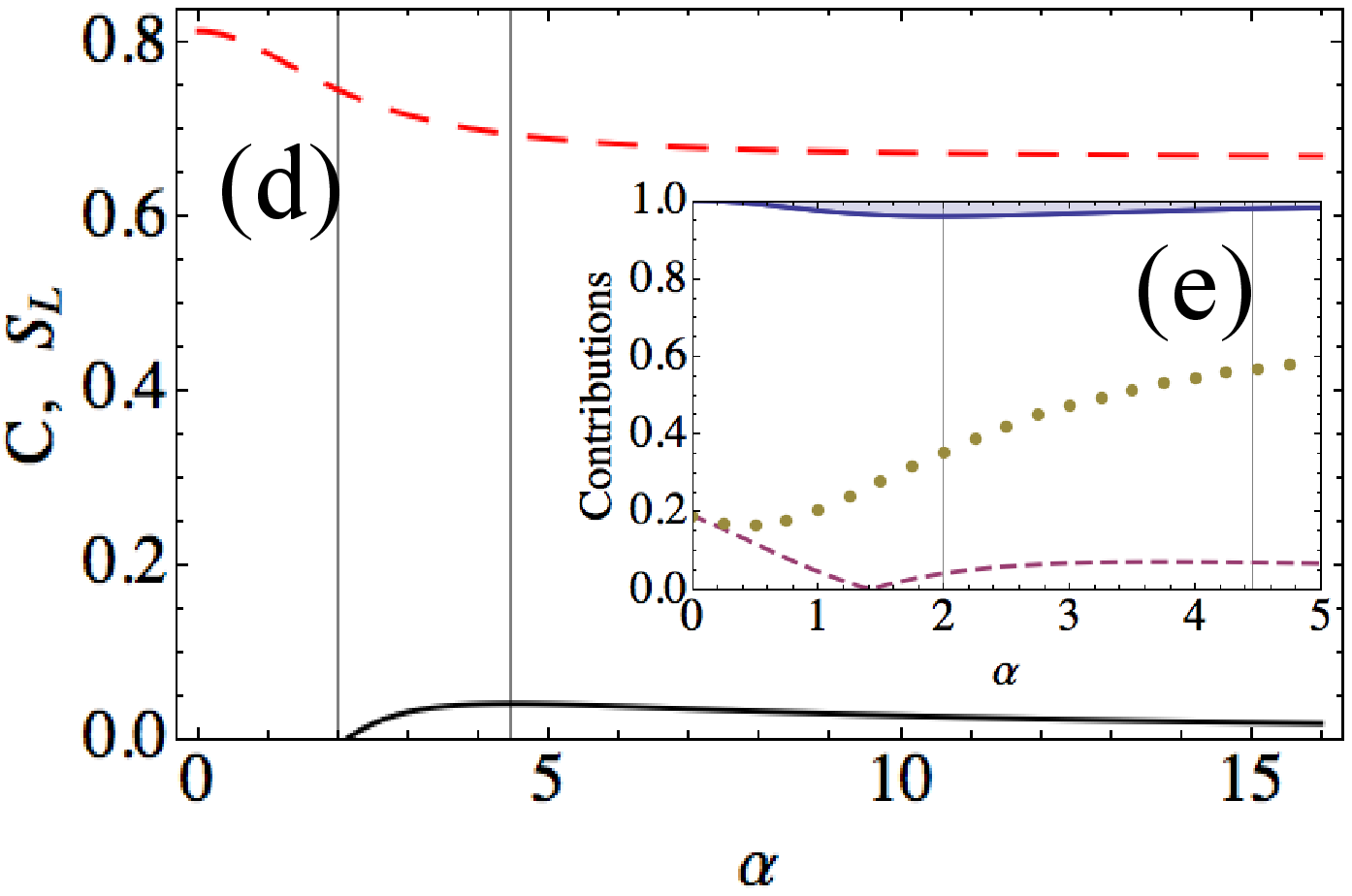}
\caption{(a) Distribution in the $C$-$S_L$ plane of all the possible
  two qubit configurations (shaded region). The thin solid line
  corresponds to the maximum $C$ for a given $S_L$ in a general
  bipartite system. The dashed blue line corresponds to the optimal
  configuration ($r_1=1$, $r_2=0$, $\Gamma_1=\Gamma_2=\Gamma$, see
  Eq.~(\ref{eq:MoFeb22170359WET2010})), a good approximation to the
  maximal $C$ vs $S_L$ in our system (with the exception of the dark
  blue region above). Below, in dark purple, the particular case of
  thermal baths. (b) $C$ (solid black) and $S_L$ (dashed red) for the
  optimal case as a function of
  $\alpha=\sqrt{\Delta^2+\Gamma^2}/g$. In inset (c), the non-entangled
  contributions to the steady state: $\tilde R_1$ (dotted brown),
  $\tilde R_2$ (dashed purple) and $\tilde R$ (solid blue). The shaded
  area represents $\tilde R_\psi$ (as $\tilde R_\psi+\tilde R
  =1$). Idem in (d) and (e), but for two thermal baths at infinite and
  zero temperatures: $r_1=1/2$, $r_2=0$. The vertical guide lines mark
  the points where entanglement appears and where it is maximum, close
  to the points where $\tilde R$ is minimum and $\tilde R$ approaches
  $\tilde R_1$, respectively.}
\label{fig:ThuFeb11135831WET2010}
\end{figure}

We start by noting that if $r_1=r_2=r$, that is, if the reservoirs are
of the same nature, there is no entanglement in the system ($C=0$),
regardless of all the other parameters, since, in this case, the
density matrix is diagonal with elements $\{(1-r)^2,r(1-r),r(1-r),
r^2\}$, that is, a mixture of separable states. This result has
already been pointed out in the literature~\cite{xujb05a,hartmann07a},
however, let us insist on the fact that, it is not the amount of
decoherence induced on the qubits by their environments what destroys
entanglement, but their similarity in nature (or temperatures in the
case of thermal baths). Let us then consider the cases with $r_2 <
r_1$ in the rest of this section.

As in Ref.~\cite{munro01a}, I examine the region of the
concurrence-linear entropy plane that our system can access in
Fig.~\ref{fig:ThuFeb11135831WET2010}(a). The shaded region is
reconstructed by randomly choosing all the parameters and computing
their $C$ and $S_L$. The accessible region is well below the black
thin line for the \emph{maximally entangled mixed
  states}~\cite{munro01a}, that provides the maximum concurrence
achievable for a given linear entropy. More interestingly, the points
are bounded in good approximation by a second (dashed blue) line
specific to our system. This line corresponds to the extreme case of
reservoirs with exactly opposite natures, $r_1=1$ and $r_2=0$, but
equally strong influence on the qubits, $\Gamma_1=\Gamma_2=\Gamma$
(that is, $P_1=\gamma_2=\Gamma$, $P_2=\gamma_1=0$). The steady state
can be written in terms of a single unit-less complex number, 
\begin{equation}
  \label{eq:SunMay23141100BST2010}
\alpha e^{i\beta}\equiv (\Delta-i\Gamma)/g\,,
\end{equation}
with norm $\alpha$ and phase $\beta$:
$\rho_{00}=\rho_{22}=\rho_{33}=\frac{1}{4+\alpha^2}$,
$\rho_{11}=\frac{1+\alpha^2}{4+\alpha^2}$ and $\rho_{12}=\frac{\alpha
  e^{-i\beta}}{4+\alpha^2}$. The two qubits are sharing a single
excitation $\mean{n_1}+\mean{n_2}=1$. Note that both detuning
($\Delta$) and the average decoherence ($\Gamma$) contribute
symmetrically to $\alpha$ and have the same effect on the steady
state: to make the coherent coupling less effective. The phase
$\beta=-\arctan{(\Gamma/\Delta)}$, which is the same than that of the
entangled state $\ket{\psi}$ formed in the steady state, can be
rotated by changing these two parameters. This is a way to phase shift
the entangled state obtained in the system. Concurrence and linear
entropy read
\begin{subequations}
    \label{eq:ThuFeb11170132WET2010}
\begin{align}
  &C=2\mathrm{Max}[\{0,\frac{\alpha-1}{4+\alpha^2}\}]\nonumber \\
  &\phantom{C}=\mathrm{Max}[\{0,\frac{\sqrt{\frac{3S_L}{2}(1-\frac{3S_L}{2}+\sqrt{1-\frac{3S_L}{4}})}-\frac{3S_L}{4}}{1+\sqrt{1-\frac{3S_L}{4}}}\}]\,,\label{eq:MoFeb22170359WET2010}\\
  &S_L=\frac{16}{3}\frac{3+\alpha^2}{(4+\alpha^2)^2}\label{eq:TueFeb23175323WET2010}\,.
\end{align}
\end{subequations}
I plot them in Fig.~\ref{fig:ThuFeb11135831WET2010}(b), as a function
of $\alpha$ moving leftwards along the dashed blue line of
Fig.~\ref{fig:ThuFeb11135831WET2010}(a). The contributions $\tilde
R_1$ (dotted brown), $\tilde R_2$ (dashed purple), $\tilde R$ (solid
blue) and $\tilde R_\psi$ (shaded area) are also presented in the
inset, Fig.~\ref{fig:ThuFeb11135831WET2010}(c), for a better
understanding of the origin of entanglement. At vanishing $\alpha$ (or
large effective coupling), the excitation is equally shared among the
states and the system is maximally mixed (with $\tilde
R_\psi=0$). Concurrence becomes different from zero at $\alpha=1$,
which is close to the point where the non-entangled contribution to
the density matrix, $\tilde R$ reaches its minimum (and $\tilde
R_\psi$ its maximum). The contributions of the spurious states
$\rho_{00}$, $\rho_{33}$ have been considerably reduced while the
coherence $|\rho_{12}|$ is sufficiently large to overcome them.

The maximum concurrence in the absolute for this system,
\begin{equation}
  \label{eq:SunMay23170206BST2010}
  C_\mathrm{max}=(\sqrt{5}-1)/4\approx 31\%\,,
\end{equation}
is reached at $\alpha=1+\sqrt{5}$. This is the region where a large
contribution $R_\psi$ is combined with low $S_L$. Moreover, the
non-entangled contribution $\tilde R$ becomes similar to $\tilde R_1$
meaning that the steady state is close to the mixture
$M_{\psi}=R_{\psi}\ket{\psi}\bra{\psi}+(1-R_{\psi})\ket{1}\bra{1}$,
only between the entangled state $\ket{\psi}$ and $\ket{1}$. The large
contribution of $\ket{1}$ to the steady state is expected since the
first qubit is pumped and the second decays. What is less expected is
that, by populating this state, we are purifying the total mixture and
enhancing the presence of the entangled state. Increasing $\alpha$
further leads to the saturation of the system into state $M_{\psi}$
and eventually to \emph{self-quenching} of
coherence~\cite{benson99a}. Note, however, that concurrence decreases
slowly and never becomes strictly zero again, due to the fact that
$\rho\rightarrow M_{\psi}$ and, therefore, $C\rightarrow R_{\psi}$.

\begin{figure}[t]
\centering
\includegraphics[width=\linewidth]{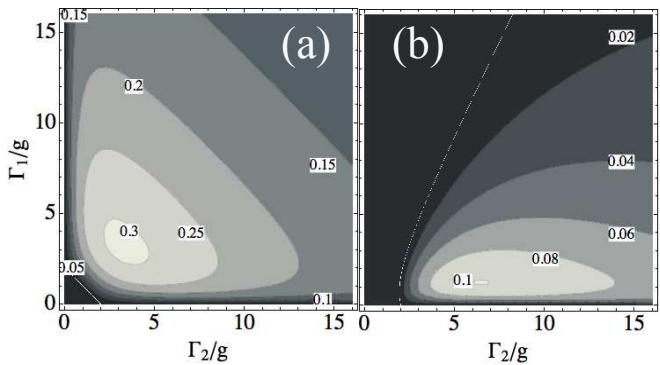}
\caption{Contour plots of concurrence $C$ as a function of $\Gamma_1$
  and $\Gamma_2$ for (a) $r_1=1$, $r_2=0$ and (b) $r_1=1/2$, $r_2=0$.
  To the left of the white lines, $C=0$. The maximum value achieved
  with this system is in (a), $C_\mathrm{max}\approx 31\%$. In (b),
  with thermal baths, concurrence rises up to $C\approx 10\%$ for an
  asymmetric configuration.}
\label{fig:MoMar8163622WET2010}
\end{figure}

The small region in dark blue in
Fig.~\ref{fig:ThuFeb11135831WET2010}(a), to the right and above the
line in Eq.~(\ref{eq:MoFeb22170359WET2010}), corresponds to cases more
entangled for the same entropy, than the configuration previously
discussed. Relaxing the previous conditions to $\Gamma_1\neq\Gamma_2$,
for instance, is enough to fill this area. In any case, configurations
above the dashed line exist only for very mixed states, with
$S_L>(17-3\sqrt{5})/30\approx 0.34$, being less appealing for
applications. In Fig.~\ref{fig:MoMar8163622WET2010}(a), I plot the
corresponding concurrence as a function of $\Gamma_1$ and $\Gamma_2$
in order to show that it is robust to their difference: $C>0$ as long
as $\Gamma_1+\Gamma_2>2$, and $C>C_\mathrm{max}/2\approx 15\%$ in most
of the area shown.

To conclude this analysis, one can check that if one medium provides
an overall dissipation and the other one, an overall gain
($0<r_2<1/2<r_1<1$), then $C$ can reach non-negligible values (above
$10\%$). This is one of the important results in this text, the
opposite nature of the reservoirs can lead the steady state close to
$M_{\psi}$, allowing for the highest degrees of entanglement in the
system.

On the other hand, if we keep equal interactions with the reservoirs,
$\Gamma_1=\Gamma_2=\Gamma$, but they are not restricted in their
natures, the values of the concurrence decreases. Let us consider the
case that has been previously studied in the literature, two thermal
reservoirs in contact with two qubits. In our notation:
$0<r_2<r_1<1/2$. In this case, the concurrence does not grow higher
than $4\%$ (as we said in the introduction), which is reached for the
extreme case, $r_1=1/2$, $r_2=0$. Again, within the new restrictions,
it is favorable for entanglement that excitation is provided through
one qubit while the other only dissipates. The equivalent expressions
to Eqs.~(\ref{eq:ThuFeb11170132WET2010}) read:
\begin{subequations}
    \label{ThuMar18181726WET2010}
  \begin{align}
    &C=\mathrm{Max}[\{0,\frac{\alpha-\sqrt{9/4+\alpha^2/2}}{4+\alpha^2}\}]\,,\\
    &S_L=\frac{39+2\alpha^2(9+\alpha^2)}{3(4+\alpha^2)^2}\,.
  \end{align}
\end{subequations}
This case is featured in
Fig.~\ref{fig:ThuFeb11135831WET2010}(d)-(e). We observe that $C$
becomes different from zero again at the minimal $\tilde R$ (maximal
entangled contribution $\tilde R_\psi$) and that its maximum value is
reached when $\tilde R_1$ approaches $\tilde R$. However, the
concurrence remains one order of magnitude smaller than
$C_\mathrm{max}$ and the linear entropy does not drop to 0. With
thermal excitation, $\tilde R_\psi$ is always too small and the steady
state is not close enough to $M_{\psi}$ to exhibit a high degree of
entanglement. However, in contrast with the optimally pumped case, one
can increase entanglement from these figures by allowing $\Gamma_1\neq
\Gamma_2$. Concurrence is increased, filling the purple darker shaded
region in Fig.~\ref{fig:ThuFeb11135831WET2010}(a). The maximum
concurrence here is $C\approx 10\%$ at $\Gamma_1\approx 1.24$ and
$\Gamma_2\approx 6.45$. This is shown in
Fig.~\ref{fig:MoMar8163622WET2010}(b) where the highest values of $C$
appear in light grey around those rates.

\section{Antibunching}
\label{sec:MonMay17124051BST2010}

Is there an experimental observable that can evidence the high degrees
of entanglement that we have analysed here? One possibility is to
reconstruct the steady state density matrix through quantum
tomography, but this method involves complicated set ups and numerous
and repeated measurements~\cite{steffen06a}. Here, I propose an
alternative method that only involves photon counting, that is, the
quantity:
\begin{equation}
  \label{eq:2309v092e}
  \delta\equiv \mean{n_1} \mean{n_2}-\mean{n_1 n_2}=\rho_{11}\rho_{22}-\rho_{00}\rho_{33}\,.
\end{equation}
$\mean{n_1}$ and $\mean{n_2}$ are proportional to the intensity of the
light emitted from each qubit, obtained by counting photons from each
source, while $\mean{n_1 n_2}$ is obtained by counting simultaneous
photon detections. $\delta$ is directly linked to the second order
cross correlation function~\cite{delvalle10b} at zero delay,
$g_{12}^{(2)}(0)=1-\delta/(\mean{n_1} \mean{n_2})$. $\delta$ is zero
if the qubits, acting as two random photon sources, are independent
($g_{12}^{(2)}(0)=g_{12}^{(2)}(\infty)=1$), and different from zero if
one qubit's emission is conditional to the other qubit's
state. $\delta<0$ implies that the simultaneous emission from both
qubits is enhanced in the system as compared to the independent
emissions. This is a necessary condition for photon \emph{bunching},
although bunching also requires
$g_{12}^{(2)}(0)>g_{12}^{(2)}(\tau)$. An example where $\delta<0$, is
the cross simultaneous emission of two coupled harmonic
oscillators~\cite{delvalle_book10a}. On the other hand, $\delta>0$
implies that simultaneous emission from both qubits is less likely
than in the uncoupled situation. Again, this is necessary for photon
\emph{antibunching} ($g_{12}^{(2)}(0)<g_{12}^{(2)}(\tau)$).

\begin{figure}[t]
\centering
\includegraphics[width=\linewidth]{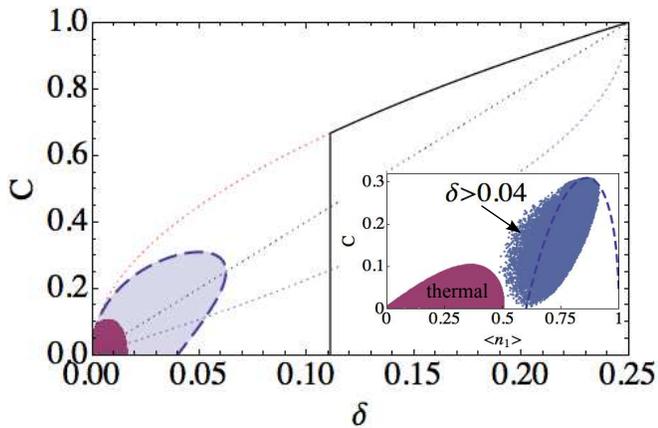}
\caption{Distribution in the $C$-$\delta$ plane of all the possible
  qubit configurations (shaded region). The thin solid and dotted
  lines correspond to different examples of entangled mixed states,
  with $\ket{\psi}$ (see the main text). The pure state $\ket{\psi}$
  corresponds to the extreme point $(1/4,1)$. The dashed blue line
  corresponds to the configuration ($r_1=1$, $r_2=0$,
  $\Gamma_1=\Gamma_2=\Gamma$, increasing $\Gamma$ anti-clockwise)
  which encloses all possible realizations. Inside, in dark purple,
  the particular case of thermal reservoirs. In inset: $C$ vs
  $\mean{n_1}$ for thermal reservoirs (where $\mean{n_1}<0.5$, in dark
  purple) and for those configurations where $\delta>0.04$ (in
  blue). The dashed blue line ($r_1=1$, $r_2=0$,
  $\Gamma_1=\Gamma_2=\Gamma$) goes clockwise with increasing
  $\Gamma$.}
\label{fig:SatMay8145927BST2010}
\end{figure}

In the steady state of our system, the emission from one of the qubits
is always antibunched ($g_{ii}^{(2)}(0)=0<g_{ii}^{(2)}(\tau)$,
$i=1,2$, as it correspond to a two-level system) and the cross
emission from both qubits fits $0\leq \delta\leq 1/4$. One can check
these limits from gathering $\delta$s from many randomly generated
configurations. It cannot go below zero because the only coherence and
entanglement in the system come from the state~$\ket{\psi}$ (that
gives the maximum value $\delta=1/4$) and not~$\ket{\phi}\equiv
(\ket{0}+e^{i\beta'}\ket{3})/\sqrt{2}$ (that would give
$\delta=-1/4$). The sign of $\delta$ is linked to the type of
entangled state realised in the system, also when there is
superposition or mixture with other states and $C<1$. For instance, if
we plot $C$ versus $\delta$ for the maximally entangled mixed
states~\cite{munro01a} with entanglement provided by $\ket{\psi}$, we
obtain the black thin line in Fig.~\ref{fig:SatMay8145927BST2010},
that abruptly falls at $\delta=1/9$. This is because for this kind of
states with $C<2/3$, $\delta$ remains constant. One would obtain the
symmetrical curve at negative $\delta$, if the mixture was with
$\ket{\phi}$. In this example, high degrees of entanglement are
related to large $\delta$. The other dotted lines appearing in
Fig.~\ref{fig:SatMay8145927BST2010} are more examples of this
relationship between the type of entanglement and $C$ with
$\delta$. The central black dotted line corresponds to the
superposition or a mixture of $\ket{\psi}$ with $\ket{\phi}$, when
$\ket{\psi}$ is the dominant state: $C=4\delta$. There is a symmetric
counterpart curve (not shown) in the opposite situation, where
$\ket{\phi}$ is dominant, with $C=4|\delta|$ and $\delta<0$. The upper
red dotted line corresponds to the superposition or mixture of
$\ket{\psi}$ with $\ket{0}$ or with $\ket{3}$: $C=2\sqrt{\delta}$. The
counterpart curve, with $\ket{\phi}$, is symmetrical. The space
between these two dotted lines could be filled with mixtures of
$\ket{\psi}$ with both $\ket{0}$ and $\ket{3}$. The lower blue dotted
line corresponds to the mixture of $\ket{\psi}$ with $\ket{1}$, the
state $M_\psi$: $C=1-\sqrt{1-4\delta}$. In all these cases, large
$\delta$ is correlated with large $C$, although the connexion is
rather general and not exclusive enough to define any
\emph{entanglement witness} in terms of $\delta$.

Let us go back to our system and investigate how to use these
correlations to extract information about $C$ from the measured
$\delta$. The shaded region in Fig.~\ref{fig:SatMay8145927BST2010}
corresponds, as in Fig.~\ref{fig:ThuFeb11135831WET2010}, to the
situations realised in our system. It is completely enclosed this time
by the dashed blue line, which corresponds to reservoirs with opposite
natures (as analysed in the previous section). In this limiting case,
$\delta$ reads
\begin{equation}
  \label{eq:SatMay8191459BST2010}
  \delta=\Big(\frac{\alpha}{4+\alpha^2}\Big)^2\,.
\end{equation}
Thanks to this analytical boundary, we can turn the general statement
that there is some correlation between $\delta$ and $C$ into a more
accurate (mathematical) one: $C_-(\delta)\leq C \leq C_+(\delta)$
where
\begin{equation}
  \label{eq:SunMay9120739BST2010}
  C_\pm(\delta)=\frac{2\sqrt{2\delta}\sqrt{1\pm\sqrt{1-16\delta}-8\delta}-\sqrt{2\delta}}{1\pm\sqrt{1-16\delta}}\,.
\end{equation}
These inequalities become most stringent when $\delta>0.04$, for
instance $\delta>0.061$ implies $20\%<C<28.3\%$. More precise
information can be obtained if $\mean{n_1}$ is included in the
analysis, looking at the inset of
Fig.~\ref{fig:SatMay8145927BST2010}. In blue, we see a cloud of
numerically generated points where $\delta>0.04$. There is also a
clear correlation between large, unsaturated, population of the dot
($0.8<\mean{n_1}<0.91$) and large degrees of entanglement
($10\%<C<C_\mathrm{max}$).

Such large $\delta$ and $C$, cannot be obtained with thermal
reservoirs for the qubits. The small accessible area in that case, is
shaded in darker purple in Fig.~\ref{fig:SatMay8145927BST2010} (within
the dashed blue curve) and in the inset. One cannot use $\delta$ or
$\mean{n_1}$ as entanglement indicators because, for all $\delta$ or
$\mean{n_1}$, $C$ can take a broad range of values that always
includes 0.

\section{Pure dephasing}
\label{sec:MonMay17124101BST2010}

\begin{figure}[t]
\centering
\includegraphics[width=0.49\linewidth]{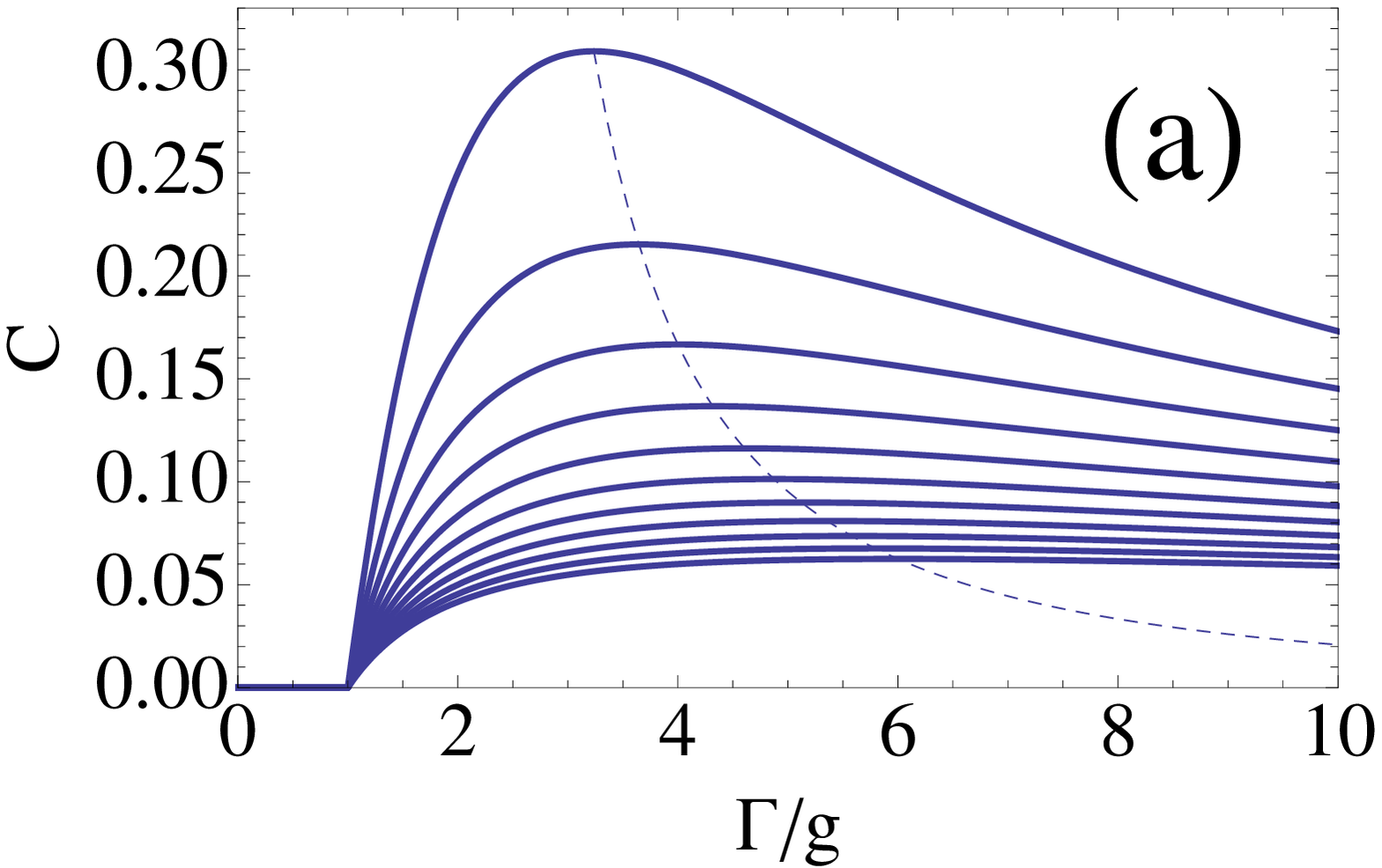}
\includegraphics[width=0.49\linewidth]{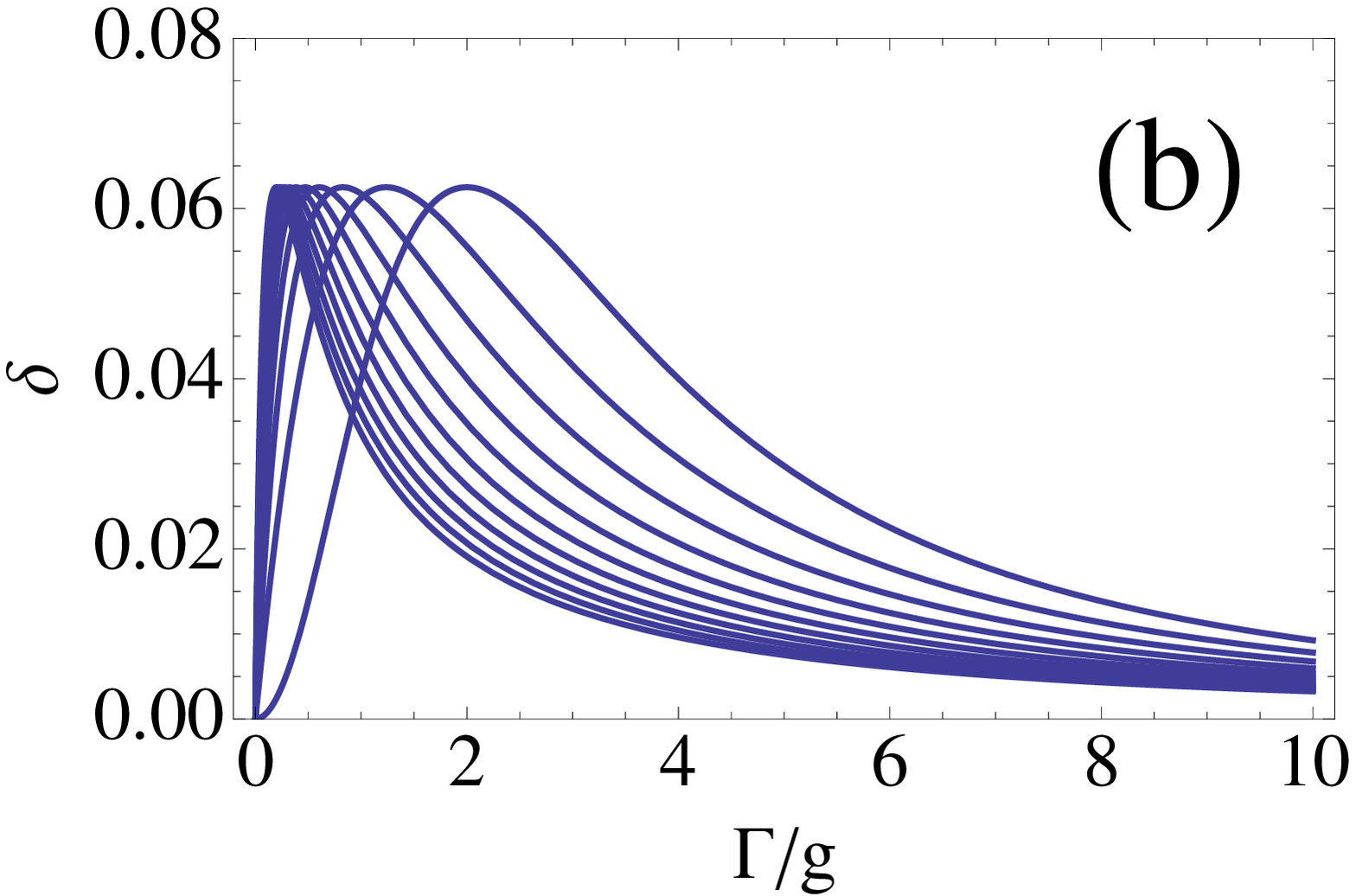}
\caption{Effect of dephasing on the results for opposite reservoirs
  ($r_1=1$, $r_2=0$, $\Gamma_1=\Gamma_2=\Gamma$), of $C$ (a) and
  $\delta$ (b) as a function of $\Gamma$. The set of curves
  corresponds to values of $\gamma^d$ from 0 to 20$g$, increasing in
  steps of $2g$. Entanglement (a) is diminished by pure dephasing,
  going from top to bottom curves, and the maximum is reached at
  higher $\Gamma$s (see the dashed line, joining the maxima of all
  curves). The maximum $\delta$ remains $1/16$ for all values of
  dephasing although this is reached for lower $\Gamma$s.}
\label{fig:SunMay9235914BST2010}
\end{figure}

Pure dephasing provides extra decay for the coherence in the
system. It weakens the correlations established between the qubits and
is, therefore, an enemy of entanglement. We can see this in
Fig.~\ref{fig:SunMay9235914BST2010}(a) where I plot the effect of
increasing dephasing ($\gamma_1^d=\gamma_2^d=\gamma^d$) on
entanglement. The curve at the top is the same as in
Fig.~\ref{fig:ThuFeb11135831WET2010}(b), with opposite kinds of
reservoirs. The rest of the curves correspond to increasing values of
the dephasing rate in steps of $2g$ up to $\gamma^d=20g$. Entanglement
decreases, and its maximum value for a given $\gamma^d$, requires
higher $\Gamma$. The set of maximum $C$ and the corresponding required
$\Gamma$ are plotted with a dashed thin line superimposed to the
curves for clarity. Entanglement is quite robust in this
configuration, it disappears but asymptotically and very slowly. Note
that $\gamma^d$ must be one order of magnitude larger than $g$ so that
$C$ is decreased to the values obtained with thermal reservoirs
($10\%$).

In Fig.~\ref{fig:SunMay9235914BST2010}(b), I plot the counterpart
curves for $\delta$, that shrink and move leftwards with
dephasing. However, the maximum $\delta$ remains $1/16$ for all
dephasing, taking place at lower $\Gamma$s. Given that the tendency of
the maximum $\delta$ is the opposite to that of the maximum $C$, the
possibility of using $\delta$ as an indicator of entanglement fails at
large dephasing.

\section{Conclusions}
\label{sec:MonMay17124110BST2010}

I have computed the entanglement ($C$) and linear entropy ($S_L$) for
two coupled qubits in the steady state created by an incoherent
continuous excitation. I have studied, not only the case where the
excitation is of a thermal origin, but also a more general
out-of-equilibrium situation where populations can be inverted
($\mean{n_1}>0.5$). In this case, I find that entanglement can be
greatly enhanced ($C$ up to $31\%$) as compared to the best thermal
values ($C$ up to $10\%$), with also a much higher purity of the
state. This is obtained in a configuration where one qubit essentially
dissipates excitation while the other essentially gains it. I have
used both numerical results, in the most general case, and analytical
formula, in this optimal case, to fully understand and characterise
entanglement formation. Entanglement (provided by the entangled state
$\ket{\psi}$) is enhanced in the steady state when the pumped qubit
approaches saturation, because this removes population from spurious
states ($\ket{0}$ and $\ket{3}$) and purifies the statistical mixture,
even in the presence of pure dephasing. Finally, I have shown that the
quantity $\delta=\mean{n_1}\mean{n_2}-\mean{n_1n_2}$, that can be
measured experimentally by photon counting, can be used as an
indicator of high degrees of entanglement in this system and specially
for the optimal configuration.

\textbf{Acknowledgments:} This research was supported by the Newton
International Fellowship program.

\bibliography{qubits,Sci,books}

\begin{thebibliography}{53}
\expandafter\ifx\csname natexlab\endcsname\relax\def\natexlab#1{#1}\fi
\expandafter\ifx\csname bibnamefont\endcsname\relax
  \def\bibnamefont#1{#1}\fi
\expandafter\ifx\csname bibfnamefont\endcsname\relax
  \def\bibfnamefont#1{#1}\fi
\expandafter\ifx\csname citenamefont\endcsname\relax
  \def\citenamefont#1{#1}\fi
\expandafter\ifx\csname url\endcsname\relax
  \def\url#1{\texttt{#1}}\fi
\expandafter\ifx\csname urlprefix\endcsname\relax\def\urlprefix{URL }\fi
\providecommand{\bibinfo}[2]{#2}
\providecommand{\eprint}[2][]{\url{#2}}

\bibitem[{\citenamefont{Horodecki et~al.}(2009)\citenamefont{Horodecki,
  Horodecki, Horodecki, and Horodecki}}]{horodecki09a}
\bibinfo{author}{\bibfnamefont{R.}~\bibnamefont{Horodecki}},
  \bibinfo{author}{\bibfnamefont{P.}~\bibnamefont{Horodecki}},
  \bibinfo{author}{\bibfnamefont{M.}~\bibnamefont{Horodecki}},
  \bibnamefont{and}
  \bibinfo{author}{\bibfnamefont{K.}~\bibnamefont{Horodecki}},
  \bibinfo{journal}{Rev. Mod. Phys.} \textbf{\bibinfo{volume}{81}},
  \bibinfo{pages}{865} (\bibinfo{year}{2009}).

\bibitem[{\citenamefont{Nielsen and Chuang}(2000)}]{nielsen_book00a}
\bibinfo{author}{\bibfnamefont{M.~A.} \bibnamefont{Nielsen}} \bibnamefont{and}
  \bibinfo{author}{\bibfnamefont{I.~L.} \bibnamefont{Chuang}},
  \emph{\bibinfo{title}{Quantum computation and quantum information}}
  (\bibinfo{publisher}{Cambridge University Press}, \bibinfo{year}{2000}).

\bibitem[{\citenamefont{Ladd et~al.}(2010)\citenamefont{Ladd, Jelezko,
  Laflamme, Nakamura, Monroe, and OÕBrien}}]{ladd10a}
\bibinfo{author}{\bibfnamefont{T.~D.} \bibnamefont{Ladd}},
  \bibinfo{author}{\bibfnamefont{F.}~\bibnamefont{Jelezko}},
  \bibinfo{author}{\bibfnamefont{R.}~\bibnamefont{Laflamme}},
  \bibinfo{author}{\bibfnamefont{Y.}~\bibnamefont{Nakamura}},
  \bibinfo{author}{\bibfnamefont{C.}~\bibnamefont{Monroe}}, \bibnamefont{and}
  \bibinfo{author}{\bibfnamefont{J.~L.} \bibnamefont{OÕBrien}},
  \bibinfo{journal}{Nature} \textbf{\bibinfo{volume}{464}}, \bibinfo{pages}{45}
  (\bibinfo{year}{2010}).

\bibitem[{\citenamefont{Gardiner and Zoller}(2000)}]{gardiner_book00a}
\bibinfo{author}{\bibfnamefont{G.~W.} \bibnamefont{Gardiner}} \bibnamefont{and}
  \bibinfo{author}{\bibfnamefont{P.}~\bibnamefont{Zoller}},
  \emph{\bibinfo{title}{Quantum Noise}} (\bibinfo{publisher}{Springer-Verlag,
  Berlin}, \bibinfo{year}{2000}), \bibinfo{edition}{2nd} ed.

\bibitem[{\citenamefont{Ficek and Tanas}(2002)}]{ficek02a}
\bibinfo{author}{\bibfnamefont{Z.}~\bibnamefont{Ficek}} \bibnamefont{and}
  \bibinfo{author}{\bibfnamefont{R.}~\bibnamefont{Tanas}},
  \bibinfo{journal}{Phys. Rep.} \textbf{\bibinfo{volume}{372}},
  \bibinfo{pages}{369} (\bibinfo{year}{2002}).

\bibitem[{\citenamefont{Verstraete et~al.}(2009)\citenamefont{Verstraete, Wolf,
  and Cirac}}]{verstraete09a}
\bibinfo{author}{\bibfnamefont{F.}~\bibnamefont{Verstraete}},
  \bibinfo{author}{\bibfnamefont{M.~M.} \bibnamefont{Wolf}}, \bibnamefont{and}
  \bibinfo{author}{\bibfnamefont{J.~I.} \bibnamefont{Cirac}},
  \bibinfo{journal}{Nature Phys.} \textbf{\bibinfo{volume}{5}},
  \bibinfo{pages}{633} (\bibinfo{year}{2009}).

\bibitem[{\citenamefont{Yu and Eberly}(2009)}]{yu09a}
\bibinfo{author}{\bibfnamefont{T.}~\bibnamefont{Yu}} \bibnamefont{and}
  \bibinfo{author}{\bibfnamefont{J.~H.} \bibnamefont{Eberly}},
  \bibinfo{journal}{Science} \textbf{\bibinfo{volume}{323}},
  \bibinfo{pages}{598} (\bibinfo{year}{2009}).

\bibitem[{\citenamefont{Ficek}(2010)}]{ficek10a}
\bibinfo{author}{\bibfnamefont{Z.}~\bibnamefont{Ficek}},
  \bibinfo{journal}{Front. Phys. China} \textbf{\bibinfo{volume}{5}},
  \bibinfo{pages}{26} (\bibinfo{year}{2010}).

\bibitem[{\citenamefont{Braun}(2002)}]{braun02a}
\bibinfo{author}{\bibfnamefont{D.}~\bibnamefont{Braun}},
  \bibinfo{journal}{Phys. Rev. Lett.} \textbf{\bibinfo{volume}{89}},
  \bibinfo{pages}{277901} (\bibinfo{year}{2002}).

\bibitem[{\citenamefont{Kim et~al.}(2002)\citenamefont{Kim, Lee, Ahn, and
  Knight}}]{kim02a}
\bibinfo{author}{\bibfnamefont{M.~S.} \bibnamefont{Kim}},
  \bibinfo{author}{\bibfnamefont{J.}~\bibnamefont{Lee}},
  \bibinfo{author}{\bibfnamefont{D.}~\bibnamefont{Ahn}}, \bibnamefont{and}
  \bibinfo{author}{\bibfnamefont{P.~L.} \bibnamefont{Knight}},
  \bibinfo{journal}{Phys. Rev. A} \textbf{\bibinfo{volume}{65}},
  \bibinfo{pages}{040101(R)} (\bibinfo{year}{2002}).

\bibitem[{\citenamefont{Jak—bczyk}(2002)}]{jakobczyk02a}
\bibinfo{author}{\bibfnamefont{L.}~\bibnamefont{Jak—bczyk}},
  \bibinfo{journal}{J. Phys. A: Math. Theor.} \textbf{\bibinfo{volume}{35}},
  \bibinfo{pages}{6383} (\bibinfo{year}{2002}).

\bibitem[{\citenamefont{Schneider and Milburn}(2002)}]{schneider02a}
\bibinfo{author}{\bibfnamefont{S.}~\bibnamefont{Schneider}} \bibnamefont{and}
  \bibinfo{author}{\bibfnamefont{G.~J.} \bibnamefont{Milburn}},
  \bibinfo{journal}{Phys. Rev. A} \textbf{\bibinfo{volume}{65}},
  \bibinfo{pages}{042107} (\bibinfo{year}{2002}).

\bibitem[{\citenamefont{Benatti et~al.}(2003)\citenamefont{Benatti, Floreanini,
  and Piani}}]{benatti03a}
\bibinfo{author}{\bibfnamefont{F.}~\bibnamefont{Benatti}},
  \bibinfo{author}{\bibfnamefont{R.}~\bibnamefont{Floreanini}},
  \bibnamefont{and} \bibinfo{author}{\bibfnamefont{M.}~\bibnamefont{Piani}},
  \bibinfo{journal}{Phys. Rev. Lett.} \textbf{\bibinfo{volume}{91}},
  \bibinfo{pages}{070402} (\bibinfo{year}{2003}).

\bibitem[{\citenamefont{Xiang-Ping et~al.}(2006)\citenamefont{Xiang-Ping,
  Mao-Fa, Xiao-Juan, and Jian-Wu}}]{xiang-ping06a}
\bibinfo{author}{\bibfnamefont{L.}~\bibnamefont{Xiang-Ping}},
  \bibinfo{author}{\bibfnamefont{F.}~\bibnamefont{Mao-Fa}},
  \bibinfo{author}{\bibfnamefont{Z.}~\bibnamefont{Xiao-Juan}},
  \bibnamefont{and} \bibinfo{author}{\bibfnamefont{C.}~\bibnamefont{Jian-Wu}},
  \bibinfo{journal}{Chinese Phys. Lett.} \textbf{\bibinfo{volume}{23}},
  \bibinfo{pages}{3138} (\bibinfo{year}{2006}).

\bibitem[{\citenamefont{del Valle et~al.}(2007{\natexlab{a}})\citenamefont{del
  Valle, Laussy, and Tejedor}}]{delvalle07a}
\bibinfo{author}{\bibfnamefont{E.}~\bibnamefont{del Valle}},
  \bibinfo{author}{\bibfnamefont{F.~P.} \bibnamefont{Laussy}},
  \bibnamefont{and} \bibinfo{author}{\bibfnamefont{C.}~\bibnamefont{Tejedor}},
  \bibinfo{journal}{Europhys. Lett.} \textbf{\bibinfo{volume}{80}},
  \bibinfo{pages}{57001} (\bibinfo{year}{2007}{\natexlab{a}}).

\bibitem[{\citenamefont{del Valle et~al.}(2007{\natexlab{b}})\citenamefont{del
  Valle, Laussy, Troiani, and Tejedor}}]{delvalle07b}
\bibinfo{author}{\bibfnamefont{E.}~\bibnamefont{del Valle}},
  \bibinfo{author}{\bibfnamefont{F.~P.} \bibnamefont{Laussy}},
  \bibinfo{author}{\bibfnamefont{F.}~\bibnamefont{Troiani}}, \bibnamefont{and}
  \bibinfo{author}{\bibfnamefont{C.}~\bibnamefont{Tejedor}},
  \bibinfo{journal}{Phys. Rev. B} \textbf{\bibinfo{volume}{76}},
  \bibinfo{pages}{235317} (\bibinfo{year}{2007}{\natexlab{b}}).

\bibitem[{\citenamefont{Contreras-Pulido and
  Aguado}(2008)}]{contreras-pulido08a}
\bibinfo{author}{\bibfnamefont{L.~D.} \bibnamefont{Contreras-Pulido}}
  \bibnamefont{and} \bibinfo{author}{\bibfnamefont{R.}~\bibnamefont{Aguado}},
  \bibinfo{journal}{Phys. Rev. B} \textbf{\bibinfo{volume}{77}},
  \bibinfo{pages}{155420} (\bibinfo{year}{2008}).

\bibitem[{\citenamefont{Hor-Meyll et~al.}(2009)\citenamefont{Hor-Meyll,
  Auyuanet, Borges, Arag{\~a}o, Huguenin, Khoury, and
  Davidovich}}]{hor-meyll09a}
\bibinfo{author}{\bibfnamefont{M.}~\bibnamefont{Hor-Meyll}},
  \bibinfo{author}{\bibfnamefont{A.}~\bibnamefont{Auyuanet}},
  \bibinfo{author}{\bibfnamefont{C.~V.~S.} \bibnamefont{Borges}},
  \bibinfo{author}{\bibfnamefont{A.}~\bibnamefont{Arag{\~a}o}},
  \bibinfo{author}{\bibfnamefont{J.~A.~O.} \bibnamefont{Huguenin}},
  \bibinfo{author}{\bibfnamefont{A.~Z.} \bibnamefont{Khoury}},
  \bibnamefont{and}
  \bibinfo{author}{\bibfnamefont{L.}~\bibnamefont{Davidovich}},
  \bibinfo{journal}{Phys. Rev. A} \textbf{\bibinfo{volume}{80}},
  \bibinfo{pages}{042327} (\bibinfo{year}{2009}).

\bibitem[{\citenamefont{Benatti et~al.}(2010)\citenamefont{Benatti, Floreanini,
  and Marzolino}}]{benatti10a}
\bibinfo{author}{\bibfnamefont{F.}~\bibnamefont{Benatti}},
  \bibinfo{author}{\bibfnamefont{R.}~\bibnamefont{Floreanini}},
  \bibnamefont{and}
  \bibinfo{author}{\bibfnamefont{U.}~\bibnamefont{Marzolino}},
  \bibinfo{journal}{Phys. Rev. A} \textbf{\bibinfo{volume}{81}},
  \bibinfo{pages}{012105} (\bibinfo{year}{2010}).

\bibitem[{\citenamefont{Jak—bczyk}(2010)}]{jakobczyk10a}
\bibinfo{author}{\bibfnamefont{L.}~\bibnamefont{Jak—bczyk}},
  \bibinfo{journal}{J. Phys. B: At. Mol. Opt. Phys.}
  \textbf{\bibinfo{volume}{43}}, \bibinfo{pages}{015502}
  (\bibinfo{year}{2010}).

\bibitem[{\citenamefont{Yi et~al.}(2003)\citenamefont{Yi, Yu, Zhou, and
  Song}}]{yi03a}
\bibinfo{author}{\bibfnamefont{X.~X.} \bibnamefont{Yi}},
  \bibinfo{author}{\bibfnamefont{C.~S.} \bibnamefont{Yu}},
  \bibinfo{author}{\bibfnamefont{L.}~\bibnamefont{Zhou}}, \bibnamefont{and}
  \bibinfo{author}{\bibfnamefont{H.~S.} \bibnamefont{Song}},
  \bibinfo{journal}{Phys. Rev. A} \textbf{\bibinfo{volume}{68}},
  \bibinfo{pages}{052304} (\bibinfo{year}{2003}).

\bibitem[{\citenamefont{Plenio and Huelga}(2002)}]{plenio02a}
\bibinfo{author}{\bibfnamefont{M.~B.} \bibnamefont{Plenio}} \bibnamefont{and}
  \bibinfo{author}{\bibfnamefont{S.~F.} \bibnamefont{Huelga}},
  \bibinfo{journal}{Phys. Rev. Lett.} \textbf{\bibinfo{volume}{88}},
  \bibinfo{pages}{197901} (\bibinfo{year}{2002}).

\bibitem[{\citenamefont{Hao and Zhub}(2007)}]{hao07a}
\bibinfo{author}{\bibfnamefont{X.}~\bibnamefont{Hao}} \bibnamefont{and}
  \bibinfo{author}{\bibfnamefont{S.}~\bibnamefont{Zhub}},
  \bibinfo{journal}{Eur. Phys. J. D} \textbf{\bibinfo{volume}{41}},
  \bibinfo{pages}{199} (\bibinfo{year}{2007}).

\bibitem[{\citenamefont{Huang et~al.}(2009)\citenamefont{Huang, Guo, and
  Yi}}]{huang09a}
\bibinfo{author}{\bibfnamefont{X.~L.} \bibnamefont{Huang}},
  \bibinfo{author}{\bibfnamefont{J.~L.} \bibnamefont{Guo}}, \bibnamefont{and}
  \bibinfo{author}{\bibfnamefont{X.~X.} \bibnamefont{Yi}},
  \bibinfo{journal}{Phys. Rev. A} \textbf{\bibinfo{volume}{80}},
  \bibinfo{pages}{054301} (\bibinfo{year}{2009}).

\bibitem[{\citenamefont{Xu and Li}(2005)}]{xujb05a}
\bibinfo{author}{\bibfnamefont{J.-B.} \bibnamefont{Xu}} \bibnamefont{and}
  \bibinfo{author}{\bibfnamefont{S.-B.} \bibnamefont{Li}},
  \bibinfo{journal}{New J. Phys.} \textbf{\bibinfo{volume}{7}},
  \bibinfo{pages}{72} (\bibinfo{year}{2005}).

\bibitem[{\citenamefont{Hartmann et~al.}(2006)\citenamefont{Hartmann, D{\"u}r,
  and Briegel}}]{hartmann06a}
\bibinfo{author}{\bibfnamefont{L.}~\bibnamefont{Hartmann}},
  \bibinfo{author}{\bibfnamefont{W.}~\bibnamefont{D{\"u}r}}, \bibnamefont{and}
  \bibinfo{author}{\bibfnamefont{H.-J.} \bibnamefont{Briegel}},
  \bibinfo{journal}{Phys. Rev. A} \textbf{\bibinfo{volume}{74}},
  \bibinfo{pages}{052304} (\bibinfo{year}{2006}).

\bibitem[{\citenamefont{Hartmann et~al.}(2007)\citenamefont{Hartmann, D{\"u}r,
  and Briegel}}]{hartmann07a}
\bibinfo{author}{\bibfnamefont{L.}~\bibnamefont{Hartmann}},
  \bibinfo{author}{\bibfnamefont{W.}~\bibnamefont{D{\"u}r}}, \bibnamefont{and}
  \bibinfo{author}{\bibfnamefont{H.~J.} \bibnamefont{Briegel}},
  \bibinfo{journal}{New J. Phys.} \textbf{\bibinfo{volume}{9}},
  \bibinfo{pages}{230} (\bibinfo{year}{2007}).

\bibitem[{\citenamefont{Lambert et~al.}(2007)\citenamefont{Lambert, Aguado, and
  Brandes}}]{lambert07a}
\bibinfo{author}{\bibfnamefont{N.}~\bibnamefont{Lambert}},
  \bibinfo{author}{\bibfnamefont{R.}~\bibnamefont{Aguado}}, \bibnamefont{and}
  \bibinfo{author}{\bibfnamefont{T.}~\bibnamefont{Brandes}},
  \bibinfo{journal}{Phys. Rev. B} \textbf{\bibinfo{volume}{75}},
  \bibinfo{pages}{045340} (\bibinfo{year}{2007}).

\bibitem[{\citenamefont{Rivas et~al.}(2009)\citenamefont{Rivas, Oxtoby, and
  Huelga}}]{rivas09a}
\bibinfo{author}{\bibfnamefont{A.}~\bibnamefont{Rivas}},
  \bibinfo{author}{\bibfnamefont{N.~P.} \bibnamefont{Oxtoby}},
  \bibnamefont{and} \bibinfo{author}{\bibfnamefont{S.~F.}
  \bibnamefont{Huelga}}, \bibinfo{journal}{Eur. Phys. J. B}
  \textbf{\bibinfo{volume}{69}}, \bibinfo{pages}{51} (\bibinfo{year}{2009}).

\bibitem[{\citenamefont{Wang et~al.}(2009)\citenamefont{Wang, Liu, and
  He}}]{wang09a}
\bibinfo{author}{\bibfnamefont{H.}~\bibnamefont{Wang}},
  \bibinfo{author}{\bibfnamefont{S.}~\bibnamefont{Liu}}, \bibnamefont{and}
  \bibinfo{author}{\bibfnamefont{J.}~\bibnamefont{He}}, \bibinfo{journal}{Phys.
  Rev. E} \textbf{\bibinfo{volume}{79}}, \bibinfo{pages}{041113}
  (\bibinfo{year}{2009}).

\bibitem[{\citenamefont{Zhou et~al.}(2009)\citenamefont{Zhou, Yang, and
  Patnaik}}]{zhou09a}
\bibinfo{author}{\bibfnamefont{L.}~\bibnamefont{Zhou}},
  \bibinfo{author}{\bibfnamefont{G.~H.} \bibnamefont{Yang}}, \bibnamefont{and}
  \bibinfo{author}{\bibfnamefont{A.~K.} \bibnamefont{Patnaik}},
  \bibinfo{journal}{Phys. Rev. A} \textbf{\bibinfo{volume}{79}},
  \bibinfo{pages}{062102} (\bibinfo{year}{2009}).

\bibitem[{\citenamefont{Li and Paraoanu}(2009)}]{li09b}
\bibinfo{author}{\bibfnamefont{J.}~\bibnamefont{Li}} \bibnamefont{and}
  \bibinfo{author}{\bibfnamefont{G.~S.} \bibnamefont{Paraoanu}},
  \bibinfo{journal}{New J. Phys.} \textbf{\bibinfo{volume}{41}},
  \bibinfo{pages}{113020} (\bibinfo{year}{2009}).

\bibitem[{\citenamefont{Shan et~al.}(2010)\citenamefont{Shan, Chen, Liu, Cheng,
  Liu, Huang, and Li}}]{shan10a}
\bibinfo{author}{\bibfnamefont{C.-J.} \bibnamefont{Shan}},
  \bibinfo{author}{\bibfnamefont{T.}~\bibnamefont{Chen}},
  \bibinfo{author}{\bibfnamefont{J.-B.} \bibnamefont{Liu}},
  \bibinfo{author}{\bibfnamefont{W.-W.} \bibnamefont{Cheng}},
  \bibinfo{author}{\bibfnamefont{T.-K.} \bibnamefont{Liu}},
  \bibinfo{author}{\bibfnamefont{Y.-X.} \bibnamefont{Huang}}, \bibnamefont{and}
  \bibinfo{author}{\bibfnamefont{H.}~\bibnamefont{Li}}, \bibinfo{journal}{Int.
  J. Theor. Phys.} \textbf{\bibinfo{volume}{49}}, \bibinfo{pages}{717}
  (\bibinfo{year}{2010}).

\bibitem[{\citenamefont{Bloch}(2008)}]{bloch08b}
\bibinfo{author}{\bibfnamefont{I.}~\bibnamefont{Bloch}},
  \bibinfo{journal}{Nature} \textbf{\bibinfo{volume}{453}},
  \bibinfo{pages}{1016} (\bibinfo{year}{2008}).

\bibitem[{\citenamefont{Bayer et~al.}(2001)\citenamefont{Bayer, Hawrylak,
  Hinzer, Fafard, Korkusinski, Wasilewski, Stern, and Forchel}}]{bayer01b}
\bibinfo{author}{\bibfnamefont{M.}~\bibnamefont{Bayer}},
  \bibinfo{author}{\bibfnamefont{P.}~\bibnamefont{Hawrylak}},
  \bibinfo{author}{\bibfnamefont{K.}~\bibnamefont{Hinzer}},
  \bibinfo{author}{\bibfnamefont{S.}~\bibnamefont{Fafard}},
  \bibinfo{author}{\bibfnamefont{M.}~\bibnamefont{Korkusinski}},
  \bibinfo{author}{\bibfnamefont{Z.~R.} \bibnamefont{Wasilewski}},
  \bibinfo{author}{\bibfnamefont{O.}~\bibnamefont{Stern}}, \bibnamefont{and}
  \bibinfo{author}{\bibfnamefont{A.}~\bibnamefont{Forchel}},
  \bibinfo{journal}{Science} \textbf{\bibinfo{volume}{291}},
  \bibinfo{pages}{451} (\bibinfo{year}{2001}).

\bibitem[{\citenamefont{Krenner et~al.}(2005)\citenamefont{Krenner, Sabathil,
  Clark, Kress, Schuh, Bichler, Abstreiter, and Finley}}]{krenner05b}
\bibinfo{author}{\bibfnamefont{H.~J.} \bibnamefont{Krenner}},
  \bibinfo{author}{\bibfnamefont{M.}~\bibnamefont{Sabathil}},
  \bibinfo{author}{\bibfnamefont{E.~C.} \bibnamefont{Clark}},
  \bibinfo{author}{\bibfnamefont{A.}~\bibnamefont{Kress}},
  \bibinfo{author}{\bibfnamefont{D.}~\bibnamefont{Schuh}},
  \bibinfo{author}{\bibfnamefont{M.}~\bibnamefont{Bichler}},
  \bibinfo{author}{\bibfnamefont{G.}~\bibnamefont{Abstreiter}},
  \bibnamefont{and} \bibinfo{author}{\bibfnamefont{J.~J.}
  \bibnamefont{Finley}}, \bibinfo{journal}{Phys. Rev. Lett.}
  \textbf{\bibinfo{volume}{94}}, \bibinfo{pages}{057402}
  (\bibinfo{year}{2005}).

\bibitem[{\citenamefont{Gerardot et~al.}(2005)\citenamefont{Gerardot, Strauf,
  de~Dood, Bychkov, Badolato, Hennessy, Hu, Bouwmeester, and
  Petroff}}]{gerardot05a}
\bibinfo{author}{\bibfnamefont{B.~D.} \bibnamefont{Gerardot}},
  \bibinfo{author}{\bibfnamefont{S.}~\bibnamefont{Strauf}},
  \bibinfo{author}{\bibfnamefont{M.~J.~A.} \bibnamefont{de~Dood}},
  \bibinfo{author}{\bibfnamefont{A.~M.} \bibnamefont{Bychkov}},
  \bibinfo{author}{\bibfnamefont{A.}~\bibnamefont{Badolato}},
  \bibinfo{author}{\bibfnamefont{K.}~\bibnamefont{Hennessy}},
  \bibinfo{author}{\bibfnamefont{E.~L.} \bibnamefont{Hu}},
  \bibinfo{author}{\bibfnamefont{D.}~\bibnamefont{Bouwmeester}},
  \bibnamefont{and} \bibinfo{author}{\bibfnamefont{P.~M.}
  \bibnamefont{Petroff}}, \bibinfo{journal}{Phys. Rev. Lett.}
  \textbf{\bibinfo{volume}{95}}, \bibinfo{pages}{137403}
  (\bibinfo{year}{2005}).

\bibitem[{\citenamefont{Clarke and Wilhelm}(2008)}]{clarke08a}
\bibinfo{author}{\bibfnamefont{J.}~\bibnamefont{Clarke}} \bibnamefont{and}
  \bibinfo{author}{\bibfnamefont{F.~K.} \bibnamefont{Wilhelm}},
  \bibinfo{journal}{Nature} \textbf{\bibinfo{volume}{453}},
  \bibinfo{pages}{1031} (\bibinfo{year}{2008}).

\bibitem[{\citenamefont{{\u Imamo\=glu} et~al.}(1999)\citenamefont{{\u
  Imamo\=glu}, Awschalom, Burkard, DiVincenzo, Loss, Sherwin, and
  Small}}]{imamoglu99a}
\bibinfo{author}{\bibfnamefont{A.}~\bibnamefont{{\u Imamo\=glu}}},
  \bibinfo{author}{\bibfnamefont{D.~D.} \bibnamefont{Awschalom}},
  \bibinfo{author}{\bibfnamefont{G.}~\bibnamefont{Burkard}},
  \bibinfo{author}{\bibfnamefont{D.~P.} \bibnamefont{DiVincenzo}},
  \bibinfo{author}{\bibfnamefont{D.}~\bibnamefont{Loss}},
  \bibinfo{author}{\bibfnamefont{M.}~\bibnamefont{Sherwin}}, \bibnamefont{and}
  \bibinfo{author}{\bibfnamefont{A.}~\bibnamefont{Small}},
  \bibinfo{journal}{Phys. Rev. Lett.} \textbf{\bibinfo{volume}{83}},
  \bibinfo{pages}{4204} (\bibinfo{year}{1999}).

\bibitem[{\citenamefont{Zheng and Guo}(2000)}]{shi-biao00a}
\bibinfo{author}{\bibfnamefont{S.~B.} \bibnamefont{Zheng}} \bibnamefont{and}
  \bibinfo{author}{\bibfnamefont{G.~C.} \bibnamefont{Guo}},
  \bibinfo{journal}{Phys. Rev. Lett.} \textbf{\bibinfo{volume}{85}},
  \bibinfo{pages}{2392} (\bibinfo{year}{2000}).

\bibitem[{\citenamefont{Ashhab et~al.}(2008)\citenamefont{Ashhab, Niskanen,
  Harrabi, Nakamura, Picot, de~Groot, Harmans, Mooij, and Nori}}]{ashhab08a}
\bibinfo{author}{\bibfnamefont{S.}~\bibnamefont{Ashhab}},
  \bibinfo{author}{\bibfnamefont{A.~O.} \bibnamefont{Niskanen}},
  \bibinfo{author}{\bibfnamefont{K.}~\bibnamefont{Harrabi}},
  \bibinfo{author}{\bibfnamefont{Y.}~\bibnamefont{Nakamura}},
  \bibinfo{author}{\bibfnamefont{T.}~\bibnamefont{Picot}},
  \bibinfo{author}{\bibfnamefont{P.~C.} \bibnamefont{de~Groot}},
  \bibinfo{author}{\bibfnamefont{C.~J. P.~M.} \bibnamefont{Harmans}},
  \bibinfo{author}{\bibfnamefont{J.~E.} \bibnamefont{Mooij}}, \bibnamefont{and}
  \bibinfo{author}{\bibfnamefont{F.}~\bibnamefont{Nori}},
  \bibinfo{journal}{Phys. Rev. B} \textbf{\bibinfo{volume}{77}},
  \bibinfo{pages}{014510} (\bibinfo{year}{2008}).

\bibitem[{\citenamefont{Osnaghi et~al.}(2001)\citenamefont{Osnaghi, Bertet,
  Auffeves, Maioli, Brune, Raimond, and Haroche}}]{osnaghi01a}
\bibinfo{author}{\bibfnamefont{S.}~\bibnamefont{Osnaghi}},
  \bibinfo{author}{\bibfnamefont{P.}~\bibnamefont{Bertet}},
  \bibinfo{author}{\bibfnamefont{A.}~\bibnamefont{Auffeves}},
  \bibinfo{author}{\bibfnamefont{P.}~\bibnamefont{Maioli}},
  \bibinfo{author}{\bibfnamefont{M.}~\bibnamefont{Brune}},
  \bibinfo{author}{\bibfnamefont{J.~M.} \bibnamefont{Raimond}},
  \bibnamefont{and} \bibinfo{author}{\bibfnamefont{S.}~\bibnamefont{Haroche}},
  \bibinfo{journal}{Phys. Rev. Lett.} \textbf{\bibinfo{volume}{87}},
  \bibinfo{pages}{037902} (\bibinfo{year}{2001}).

\bibitem[{\citenamefont{Majer et~al.}(2007)\citenamefont{Majer, Chow, Gambetta,
  Koch, Johnson, Schreier, Frunzio, Schuster, Houck, Wallraff
  et~al.}}]{majer07a}
\bibinfo{author}{\bibfnamefont{J.}~\bibnamefont{Majer}},
  \bibinfo{author}{\bibfnamefont{J.~M.} \bibnamefont{Chow}},
  \bibinfo{author}{\bibfnamefont{J.~M.} \bibnamefont{Gambetta}},
  \bibinfo{author}{\bibfnamefont{J.}~\bibnamefont{Koch}},
  \bibinfo{author}{\bibfnamefont{B.~R.} \bibnamefont{Johnson}},
  \bibinfo{author}{\bibfnamefont{J.~A.} \bibnamefont{Schreier}},
  \bibinfo{author}{\bibfnamefont{L.}~\bibnamefont{Frunzio}},
  \bibinfo{author}{\bibfnamefont{D.~I.} \bibnamefont{Schuster}},
  \bibinfo{author}{\bibfnamefont{A.~A.} \bibnamefont{Houck}},
  \bibinfo{author}{\bibfnamefont{A.}~\bibnamefont{Wallraff}},
  \bibnamefont{et~al.}, \bibinfo{journal}{Nature}
  \textbf{\bibinfo{volume}{449}}, \bibinfo{pages}{443} (\bibinfo{year}{2007}).

\bibitem[{\citenamefont{Laucht et~al.}(2010)\citenamefont{Laucht, Hauke,
  Villas-B\^oas, Hofbauer, Kaniber, B\"ohm, and Finley}}]{arxiv_laucht10a}
\bibinfo{author}{\bibfnamefont{A.}~\bibnamefont{Laucht}},
  \bibinfo{author}{\bibfnamefont{N.}~\bibnamefont{Hauke}},
  \bibinfo{author}{\bibfnamefont{J.~M.} \bibnamefont{Villas-B\^oas}},
  \bibinfo{author}{\bibfnamefont{F.}~\bibnamefont{Hofbauer}},
  \bibinfo{author}{\bibfnamefont{M.}~\bibnamefont{Kaniber}},
  \bibinfo{author}{\bibfnamefont{G.}~\bibnamefont{B\"ohm}}, \bibnamefont{and}
  \bibinfo{author}{\bibfnamefont{J.~J.} \bibnamefont{Finley}},
  \bibinfo{journal}{arXiv:0904.4759}  (\bibinfo{year}{2010}).

\bibitem[{\citenamefont{Gallardo et~al.}(2010)\citenamefont{Gallardo, Martinez,
  Nowak, Sarkar, van~der Meulen, Calleja, Tejedor, Prieto, Granados, Taboada
  et~al.}}]{gallardo10a}
\bibinfo{author}{\bibfnamefont{E.}~\bibnamefont{Gallardo}},
  \bibinfo{author}{\bibfnamefont{L.~J.} \bibnamefont{Martinez}},
  \bibinfo{author}{\bibfnamefont{A.~K.} \bibnamefont{Nowak}},
  \bibinfo{author}{\bibfnamefont{D.}~\bibnamefont{Sarkar}},
  \bibinfo{author}{\bibfnamefont{H.~P.} \bibnamefont{van~der Meulen}},
  \bibinfo{author}{\bibfnamefont{J.~M.} \bibnamefont{Calleja}},
  \bibinfo{author}{\bibfnamefont{C.}~\bibnamefont{Tejedor}},
  \bibinfo{author}{\bibfnamefont{I.}~\bibnamefont{Prieto}},
  \bibinfo{author}{\bibfnamefont{D.}~\bibnamefont{Granados}},
  \bibinfo{author}{\bibfnamefont{A.~G.} \bibnamefont{Taboada}},
  \bibnamefont{et~al.}, \bibinfo{journal}{Phys. Rev. B}
  \textbf{\bibinfo{volume}{81}}, \bibinfo{pages}{193301}
  (\bibinfo{year}{2010}).

\bibitem[{\citenamefont{del Valle}(2010{\natexlab{a}})}]{delvalle_book10a}
\bibinfo{author}{\bibfnamefont{E.}~\bibnamefont{del Valle}},
  \emph{\bibinfo{title}{Microcavity Quantum Electrodynamics}}
  (\bibinfo{publisher}{VDM Verlag, Saarbr\"ucken, Germany},
  \bibinfo{year}{2010}{\natexlab{a}}).

\bibitem[{\citenamefont{del Valle}(2010{\natexlab{b}})}]{delvalle10b}
\bibinfo{author}{\bibfnamefont{E.}~\bibnamefont{del Valle}},
  \bibinfo{journal}{Phys. Rev. A} \textbf{\bibinfo{volume}{81}},
  \bibinfo{pages}{053811} (\bibinfo{year}{2010}{\natexlab{b}}).

\bibitem[{\citenamefont{Carmichael}(2002)}]{carmichael_book02a}
\bibinfo{author}{\bibfnamefont{H.~J.} \bibnamefont{Carmichael}},
  \emph{\bibinfo{title}{Statistical methods in quantum optics 1}}
  (\bibinfo{publisher}{Springer}, \bibinfo{year}{2002}), \bibinfo{edition}{2nd}
  ed.

\bibitem[{\citenamefont{Briegel and Englert}(1993)}]{briegel93a}
\bibinfo{author}{\bibfnamefont{H.-J.} \bibnamefont{Briegel}} \bibnamefont{and}
  \bibinfo{author}{\bibfnamefont{B.-G.} \bibnamefont{Englert}},
  \bibinfo{journal}{Phys. Rev. A} \textbf{\bibinfo{volume}{47}},
  \bibinfo{pages}{3311} (\bibinfo{year}{1993}).

\bibitem[{\citenamefont{Wootters}(1998)}]{wootters98a}
\bibinfo{author}{\bibfnamefont{W.~K.} \bibnamefont{Wootters}},
  \bibinfo{journal}{Phys. Rev. Lett.} \textbf{\bibinfo{volume}{80}},
  \bibinfo{pages}{2245} (\bibinfo{year}{1998}).

\bibitem[{\citenamefont{Munro et~al.}(2001)\citenamefont{Munro, James, White,
  and Kwiat}}]{munro01a}
\bibinfo{author}{\bibfnamefont{W.~J.} \bibnamefont{Munro}},
  \bibinfo{author}{\bibfnamefont{D.~F.~V.} \bibnamefont{James}},
  \bibinfo{author}{\bibfnamefont{A.~G.} \bibnamefont{White}}, \bibnamefont{and}
  \bibinfo{author}{\bibfnamefont{P.~G.} \bibnamefont{Kwiat}},
  \bibinfo{journal}{Phys. Rev. A} \textbf{\bibinfo{volume}{64}},
  \bibinfo{pages}{030302(R)} (\bibinfo{year}{2001}).

\bibitem[{\citenamefont{Benson and Yamamoto}(1999)}]{benson99a}
\bibinfo{author}{\bibfnamefont{O.}~\bibnamefont{Benson}} \bibnamefont{and}
  \bibinfo{author}{\bibfnamefont{Y.}~\bibnamefont{Yamamoto}},
  \bibinfo{journal}{Phys. Rev. A} \textbf{\bibinfo{volume}{59}},
  \bibinfo{pages}{4756} (\bibinfo{year}{1999}).

\bibitem[{\citenamefont{Steffen et~al.}(2006)\citenamefont{Steffen, Ansmann,
  Bialczak, Katz, Lucero, McDermott, Neeley, Weig, Cleland, and
  Martinis}}]{steffen06a}
\bibinfo{author}{\bibfnamefont{M.}~\bibnamefont{Steffen}},
  \bibinfo{author}{\bibfnamefont{M.}~\bibnamefont{Ansmann}},
  \bibinfo{author}{\bibfnamefont{R.~C.} \bibnamefont{Bialczak}},
  \bibinfo{author}{\bibfnamefont{N.}~\bibnamefont{Katz}},
  \bibinfo{author}{\bibfnamefont{E.}~\bibnamefont{Lucero}},
  \bibinfo{author}{\bibfnamefont{R.}~\bibnamefont{McDermott}},
  \bibinfo{author}{\bibfnamefont{M.}~\bibnamefont{Neeley}},
  \bibinfo{author}{\bibfnamefont{E.~M.} \bibnamefont{Weig}},
  \bibinfo{author}{\bibfnamefont{A.~N.} \bibnamefont{Cleland}},
  \bibnamefont{and} \bibinfo{author}{\bibfnamefont{J.~M.}
  \bibnamefont{Martinis}}, \bibinfo{journal}{Science}
  \textbf{\bibinfo{volume}{313}}, \bibinfo{pages}{1423} (\bibinfo{year}{2006}).

\end{thebibliography}

\end{document}